\documentclass[journal,10pt]{IEEEtran}

\usepackage{amsfonts}
\usepackage{amssymb}
\usepackage{stfloats}
\usepackage{cite}
\usepackage{graphicx}
\usepackage{psfrag}
\usepackage{subfigure}
\usepackage{amsmath}
\usepackage{array}
\usepackage{algorithm}
\usepackage{algorithmic}
\usepackage[affil-it]{authblk}
\usepackage[english]{babel}
\usepackage{blindtext}
\usepackage{url}
\usepackage{epstopdf}
\usepackage{flushend}
\usepackage{verbatim}
\usepackage{bm}
\usepackage{multirow}
\usepackage{booktabs}
\usepackage{caption}
\usepackage{bbm}
\usepackage{color}
\newenvironment{iarray}{\begin{IEEEeqnarray}{rCl}}{\end{IEEEeqnarray}\ignorespacesafterend}

\title{A Hard and Soft Hybrid Slicing Framework for Service Level Agreement Guarantee via Deep Reinforcement Learning}

\author{Heng Zhang$^{\dagger}$, Guangjin Pan$^{\dagger}$, Shugong Xu$^{\dagger}$,~\IEEEmembership{Fellow,~IEEE}, Shunqing Zhang$^{\dagger}$, and Zhiyuan Jiang$^{\dagger}$\\
$^{\dagger}$ School of Communication \& Information Engineering, \\
Shanghai University, Shanghai, 200444, China\\
Email:\{hengzhang, guangjin\_pan, shugong, shunqing, jiangzhiyuan\}@shu.edu.cn}

\begin{document}
\maketitle

\begin{abstract}

 Network slicing is a critical driver for guaranteeing the diverse service level agreements (SLA) in 5G and future networks. Recently, deep reinforcement learning (DRL) has been widely utilized for resource allocation in network slicing. 
 However, existing related works do not consider the performance loss associated with the initial exploration phase of DRL. This paper proposes a new performance-guaranteed slicing strategy with a soft and hard hybrid slicing setting. Mainly, a common slice setting is applied to guarantee slices’ SLA when training the neural network. Moreover, the resource of the common slice tends to precisely redistribute to slices with the training of DRL until it converges. Furthermore, experiment results confirm the effectiveness of our proposed slicing framework: the slices’ SLA of the training phase can be guaranteed, and the proposed algorithm can achieve the near-optimal performance in terms of the SLA satisfaction ratio, isolation degree and spectrum maximization after convergence. 

\end{abstract}


\section{Introduction} \label{sect:intro}
With the emergence of 5G telecommunication technology, cellular networks are envisioned to cater services to a wide variety of innovative vertical applications, such as Cellular Vehicle-to-Everything (C-V2X), augmented/virtual reality (AR/VR), with heterogeneous performance requirements including high data rates, ultra-low latency and high reliability \cite{foukas2017network}. Network slicing is recognized as a promising technique to guarantee differentiated service QoS and service level agreements (SLAs). Since it can enable multiple logical networks corresponding to different network services run on top of a common physical network infrastructure such that the slices can be customized to satisfy various SLAs through virtualization, isolation techniques \cite{8685766}.

From a perspective of radio resource management, the fundamental challenge of network slicing lies in the trade-off of isolation and resource efficiency. On the one hand, to achieve non-interference between slices, the slicing system intends to ensure complete isolation between network slices. On the other hand, inherent radio spectrum scarcity promotes that all slices share a limited radio resource on-demand to ensure efficient utilization. Therefore, inter-slice radio resource allocation (IS-RRA) in the radio access network (RAN) becomes an open technical challenge \cite{ISRRA2020WC}.

In order to address the above problem, deep reinforcement learning (DRL) technology is widely applied due to its ability in model-free problems \cite{rli2018access,chen2019DRL,jie2021tcom,sun2019mix,myWCL2021}. \cite{rli2018access} investigates the application of DRL in solving radio resource slicing and priority-based core network slicing, and the results exhibit the advantage of DRL in solving model-free resource allocation problems.
Based on \cite{rli2018access}, \cite{chen2019DRL} proposed a faster convergence DRL scheme by integrating discrete normalized advantage functions (DNAF) and the deterministic policy gradient descent (DPGD) algorithm. The authors in \cite{jie2021tcom} propose a hierarchical control strategy to guarantee the long-term QoS of services and spectrum efficiency (SE), where DQN and DDPG networks are applied to solve the long-term and short-term problems, respectively. \cite{sun2019mix} and \cite{myWCL2021} develop DRL methods to heterogeneous networks (HetNets) scenarios to solve joint user association and network slicing problems. 

However, existing works for IS-RRA focus on purely hard isolation schemes where each slice is allocated with dedicated resources.
And the performance loss of such schemes caused by action exploration or network fine-tuning may be unbearable.  To minimize the performance loss during exploration phases, we propose a hard and soft hybrid slicing framework by introducing a \textit{Common slice} setting under a specific isolation degree constraint, in which UEs of all slices can utilize the resource of the common slice. Especially, the number of resources of the common slice can be significant in the initial training phase to guarantee slices' SLA. As the network training, the resource of the common slice is gradually adjusted until the DRL network converges to an optimal state.  

Overall speaking, this paper proposes a hard and soft hybrid slicing framework to guarantee the slices' SLA and maximize the SE as much as possible under a specific isolation constraint. Compared with purely hard algorithms based on DRL, the proposed scheme is capable of guaranteeing slices' SLA all the time, even in the initial training phase. Moreover, it achieves near-optimal performance in terms of SLA satisfaction, SE and isolation.

\section{System Model}
\label{sect:sys}

\subsection{Communication Model}
We consider a typical OFDMA based downlink cellular network consisting of a single base station (BS), where there exist multiple users denoted as $\mathcal{N} = \{1,2,\cdots,N \}$. Assume that the cellular network consists of a set of network slices denoted as $\mathcal{M} =  \{1,2,\cdots,M\}$ and $\mathcal{N}_m$ denotes the UEs that belongs to slice $m$.
Radio resource is divided into Transmission Time Intervals (TTIs) denoted by $t\in \{1,2,\cdots\}$ in time domain. The bandwidth is partitioned into $W$ resource blocks (RBs). The duration of a slicing window, where the resource allocated to each slice remains constant, is called \textit{epoch}, denoted by $k\in\{1,2,\cdots\}$, and each epoch contains $T$ consecutive TTIs. Consider a equal power allocation, the SINR of user $n$ at time $t$ is given as $\gamma_{n,t} = \frac{PH_{n,t}}{WN_0}$, where $P$ is the transmit power of BS and $H_{n,t}$ is the channel gain of user $n$. $N_0$ is the power of additive white Gaussian noise.

For the traditional traffic with a large packet size, e.g. eMBB traffic, the achievable rate of the user $n$ can be directly estimated according to Shannon's capacity. For the short-sized packet transmission, such as uRLLC and MTC services, the data rate falls in the finite blocklength channel coding regime \cite{polyanskiy2010channel}. Therefore, the data rate for are modeled as \eqref{datarate},
\begin{figure*}[b]
\hrulefill
\begin{iarray}
\label{datarate}
r_{n,t} = \left \{ \begin{array}{ll}
      \bigtriangleup t\cdot W_{n,t}\log_2\left(1+\gamma_{n,t} \right), &\text{for long packets transmission}\\
      \bigtriangleup t\cdot W_{n,t} \left[{\rm log} \left( 1+\gamma_{n,t} \right) - \sqrt{\frac{C_{n,t}}{l_{n,t}}}Q^{-1}\left( \epsilon \right) {\rm log} e\right], &\text{for short packets transmission}
\end{array}
\right.
\end{iarray}
\end{figure*}
where $\bigtriangleup t$ is the time duration of one TTI and $W_{n,t}$ is the  allocated RBs to UE $n$ within $t$-th TTI. $\epsilon$ is the transmission error probability, and $Q^{-1}\left(\cdot \right)$ is the inverse of the Gaussian Q-function, and $l_{n,t}$ represents the the length of codeword block in symbols, and $C_{n,t}$ is channel dispersion, given by $C_{n,t} = 1-\frac{1}{\left( 1+\gamma_{n,t}\right)^2}$.

\subsection{SLA Model}
Generally speaking, classical QoS metrics for slices' SLA include throughput, packet latency and transmission reliability. For the throughput, it can be easily derived by aggregating the amount of data that is successfully transmitted over time. For the packet delay, a detailed queuing model of UEs' packets needs to be clarified.

In this paper, the arrival distribution of traffic is characterised by the pattern of service, and there is no prior knowledge of volatile demand. The arriving packets of UEs are cached in the BS's buffer and are delivered according to the first-come-first-serve (FCFS) policy. Assume that each UE is corresponding one data queue at BS. The packet delay consists of two parts, i.e., queuing time and transmission time, where the former is influenced by scheduling policy and the latter is decided by instantaneous data rate. 

From the perspective of the network, the packet is dropped if its delay exceeds the predefined maximum packet latency \cite{netw2020mei}. The reliability is determined by the percentage of packets that are successfully delivered. Therefore, the transmission reliability of UE $n$ is expressed as
\begin{equation}
    \theta_{n} = \text{Pr}\{ D_{n,i} \leq D_m^{max} \}, n\in\mathcal{N}_m, 
\end{equation}
where $D_{n,i}$ is the delay of the $i$-th packet of UE $n$, and $D_m^{max}$ corresponds to the maximum packet delay of UEs in slice $m$. 

For the throughput, the SLA satisfaction ratio of slice $m$ within one epoch $k$ is defined as follows
\begin{equation}
    Q_{m,k}^{rate} = \frac{1}{\left|\mathcal{N}_m \right|} \sum_{n\in \mathcal{N}_m} \text{min}\left( \frac{\sum_{t=(k-1)T+1}^{kT}r_{n,t}}{R_{m}^{th}},1\right),
\end{equation}
where $R_{m}^{th}$ is the minimum data rate requirement.

For the latency and reliability, given the the maximum packet delay $D_m^{max}$, the SLA satisfaction ration can be represented by the reliability. Therefore we have 
\begin{equation}
   Q_{m,k}^{delay} =  \frac{1}{\left|\mathcal{N}_m \right|}\sum_{n\in \mathcal{N}_m} \theta_{n}^k
\end{equation}
where $\theta_{n}^k$ represents the transmission reliability of packets of UE $n$ under maximum delay constraint. Thus, we use the throughput, latency and reliability as the QoS metrics to evaluate the SLA satisfaction in the following.

\section{Hybrid Slicing Framework and Problem Formulation }
\label{sec:proformu}
\begin{figure}
    \centering
    \includegraphics[width=0.4\textwidth]{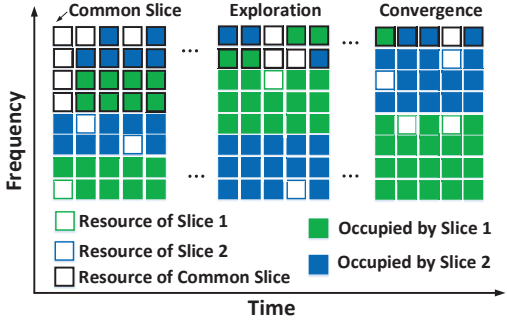}
    \caption{The illustration of the hybrid slicing framework.}
    \label{fig:hybrid_framework}
\end{figure}
\subsection{Hybrid Slicing Framework }
The purely hard slicing strategy can guarantee full isolation among slices, while it suffers from the dynamic environment and results in SLA deterioration and low resource efficiency. On the contrary, the soft slicing method can maximize resource efficiency while limited by isolation. Therefore, we propose a novel hybrid slicing framework that can take advantage of both hard and soft strategies. Especially, soft decision, i.e. common slice setting, is utilized to guarantee SLA and improve resource efficiency in the exploration phase. The hybrid slicing framework can be understood from the following two aspects.
 
 {\em 1) Common Slice Setting}: In purely hard schemes, resources dedicated to a slice need to be large enough or over-provisioning to fully guarantee the SLAs, even in the worst-case scenario of the entire slicing window. 
Fig. \ref{fig:hybrid_framework} shows a hybrid scheme, where the resources are divided into two parts, i.e. resources dedicated to slices and resources to common slice, corresponding to hard and soft strategies. All UEs can utilize the resource of the common slice according to their demand and priority. Reasonable resource configuration of the hybrid scheme enables both SLA satisfaction and resource efficiency with a small sacrifice of isolation. For example, $90\%$ resources required in worst-case scenarios can realize SLA guarantee in most cases and resources of the common slice are shared to guarantee the slices' performance of worst cases such that both SLA satisfaction and resource efficiency can be maximized under a specific isolation constraint.  

{\em 2) Periodically Adjusting Resource Slicing}: As Fig. \ref{fig:hybrid_framework} shows, radio resources can be periodically allocated to each slice to adopt a dynamic wireless environment. For example, the resources of the common slice can be significant in the initial phase to guarantee slices' SLA. As increasing awareness of the environment increases, the slice configuration converges to a precise scheme according to isolation requirement, which corresponds to the last slicing window in Fig. \ref{fig:hybrid_framework}. 
    
\subsection{Problem Formulation}
For a slice $m$, the degree of isolation in epoch $k$ is represented by follows
\begin{equation}
    o_{m,k} =  \frac{w_{m,k}}{w_{m,k}+w_{c,m,k}}
\end{equation}
where $w_{m,k}$ is the allocated resources of slice $m$ and $w_{c,m,k}$ denotes resources that slice $m$ occupies from the common slice $w_{c,k}$. The objective of the RAN slicing is to guarantee the SLA of diverse slices and simultaneously maximize the SE, which are defined as follows
\begin{iarray}
Q_{m,k} &=& f\left(d_{m,k}, r_{m,k},r_{m,c,k} \right), \label{Q_mk}\\
S_k &=& \sum_{t=\left(k-1\right)T+1}^{kT}\sum_{n=1}^{N}\frac{r_{n,t}}{W} \label{S_k}
\end{iarray}
  where $d_{m,k}$ is the fluctuation traffic demand of slice $m$, and function $f\left(\cdot\right)$ represents the complicated relationship between the SLA and traffic demand, allocated resources to slices and scheduling algorithms within slices.

 The utility function of one epoch is defined as follows
\begin{equation}
\label{utility}
    U^k = \alpha \sum_{m=1}^{M}Q_{m,k} +\beta\prod\limits_{m=1}^{M}\mathbbm{1}(Q_{m,k})\cdot S_{k},
\end{equation}
where $\alpha$ and $\beta$ are utility coefficients, and $\mathbbm{1}(Q_{m,t})$ is the indicator function to denote whether the SLA of slice $m$ is satisfied. 

The objective of a slice network is to maximize the long-term utility. A general method to maximize the average utility within a finite time period $K$, e.g., an hour, a day, or a week \cite{8931583}. Hence, the network slice problem is formulated as follows.
 \begin{eqnarray}
\mathcal{P}: \underset{w_{m,k},w_{c,k}}{\textrm{max}} && \frac{1}{K}\sum\limits_{k=1}^{K} \eqref{utility} \\
\textrm{s. t.} && \eqref{Q_mk},\eqref{S_k} \nonumber \\
&& o_{m,k} \geq o^{th}_{m}, \label{omega}\\
&& \sum_{m \in \mathcal{M}} \ w_{m,k}+w_{c,k} = W  \label{equa:w_ms},
\end{eqnarray}
where $o^{th}_{m}$ represents the threshold of required isolation. 

The difficulties of the problem $\mathcal{P}$ is reflected in two aspects. First, the heterogeneous QoS, i.e., throughput, packet delay, reliability, of slices, highly complicates the problem. Second, customized scheduling algorithms within slices and volatile traffic demand make $f\left(\cdot\right)$ extremely complex. An analytical model of $f\left(\cdot\right)$ in practical networks is almost impossible to derive \cite{myWCL2021}.
Moreover, resource allocation of slicing systems exhibit  \textit{Markovian} characteristic, i.e. the allocation strategy affects not only the current SLAs and resource efficiency but also further network state and utility, e.g., the queue of UEs and delay of packets. Therefore, DRL based solution is designed in the following section.

\section{DRL based Solution}

\subsection{Design of the DRL scheme} 
As mentioned before, the resource slicing problem can be solved by the DRL technique. In this paper, an initial slice resource allocation, e.g., NVS \cite{kokku2011nvs}, is first given. Then the DRL agent dynamically adjusts the resource allocated to slices to guarantee the SLA and isolation of slices. To achieve efficient and intelligent slicing, the agent observes the environment, e.g., performance feedback, resource utilization and so on, and makes a decision according to the observed state at the start of each epoch. The states, actions and reward of the DRL scheme is defined as follows. 

\textbf{State:} The state is defined as a tuple as follows
\begin{equation}
  s_{k} = \{w_{m,k}, Q_{m,k}, o_{m,k}, \mu_{m,k}| m \in \mathcal{M}\}
\end{equation}
where $\mu_{m,k}$ is the resource utilization of slice $m$ that is defined as the ration of used resources to the allocated resources. 

\textbf{Action:} The agent intelligently adjusts the resource allocation of slices by selecting an action $a_k$ according to the current state $s_k$. The action for a slice is defined as a set of decreasing, remaining and increasing the allocated resource. It is worth noting that the object of action interaction is the common slice. For example, slice $m$ offloads additional resources to the common slice and slice $m+1$ require more dedicated resources from the common slice at epoch $k$. And the action set of one slice is defined as $\mathcal{A}=\{-a^j, \cdots,-a^1,0,a^1,\cdots,a^j\}$, where $0<a^1<\cdots<a^{j}<W$ and $j$ is the positive integer. For example, define the action of slice $m$ is $a_{m,k}$, where $a_{m,k} \in \mathcal{A}$, we have $w_{m,k+1}=w_{m,k}+a_{m,k}$. Therefore, the action of agent at $k$ is defined as follows
\begin{equation}
    a_k = \{ a_{m,k} | m\in\mathcal{M}, a_{m,k} \in \mathcal{A}\}.
\end{equation}

\textbf{Reward:} The reward of agent is defined as follows
\begin{iarray}
r_k(s_k, a_k) = &\alpha_m& \sum_{m=1}^{M}e^{Q_{m,k}}+   \beta\prod\limits_{m=1}^{M}\mathbbm{1}(Q_{m,k})\cdot \frac{S_k}{S_{max}} \\
& -\rho&\sum_{m=1}^{M}\left[o_{m}^{th} - o_{m,k} \right]^+ \nonumber ,
\end{iarray}
where $\left[x\right]^+ =\text{max}\left(0,x\right)$, and $\rho>0$ is a punishment constant. $\frac{S_k}{S_{max}}$ operation normalizes SE by dividing the predefined maximum value $S_{max}$. The exponential reward function is to train the network more efficiently as $Q_{m,k}$ approaches 1.

\begin{figure}[!t]
    \centering
    \includegraphics[width=2.8 in]{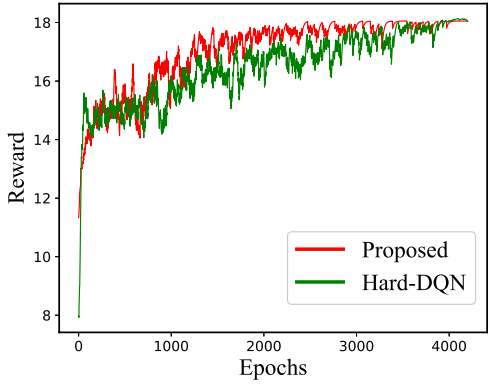}
    \caption{The convergence process of two DRL based algorithms.}
    \label{fig:train}
\end{figure}

\subsection {Training of Agents} 
A deep Q network (DQN) is applied to design and train the agent, where a neural network (NN) is used to approximate the action-value function, $q\left(s,a;\mathbf{\theta}\right) \approx Q^*\left(s,a\right)$ and $\theta$ represents the parameters of NN. The state is input to the DQN, and the network outputs the predicted Q values of each action. With the experience replay and quasistatic target network, the DQN is trained by minimizing the error between the predicted Q values and true Q values as follows,
\begin{equation}
L\left(\theta\right) = \frac{1}{B}\sum_k\left(y_k -q\left(s_k,a_k; \theta \right)\right)^2,
\end{equation}
where $B$ is the batch size. The target value $y_k$ is
\begin{equation}
y_k = r_{k+1} + \gamma\underset{a^{'}}{\text{max}}q\left(s_{k+1},a^{'};\theta^{'}\right),
\end{equation}
where $\theta^{'}$ represents the parameters of the target network and $\gamma$ is the discount factor. 

\begin{table}[b]
\centering
\caption{Slices Parameters}
\begin{tabular}{ccc}
\hline
& eMBB& uRLLC \\
\hline
Traffic Model & Poisson process & period process\\
Packet Size & 55k bits & 256 bits\\
Arrival Rate  & 100 packets/s & 100  packets/s\\
SLA & 95\% \{5M bps\} & 99\% \{5 ms and 99.99\% \} \\
$\alpha$ & 2 & 3\\
$o_m^{th}$ & 80\% & 90\% \\
Number of UEs & 20 & 50\\
Schedule & Proportional Fairness & Earliest Deadline First\\
\hline
\end{tabular}
\label{tab:slice}
\end{table}

\section{Numerical Results}
\subsection{Experiments Setup}
\begin{figure}[t!]  
\centering  \subfigure[The proposed algorithm]{
\label{iso1}
\begin{minipage}[t]{2.8in}
\centering
\includegraphics[width=2.8in]{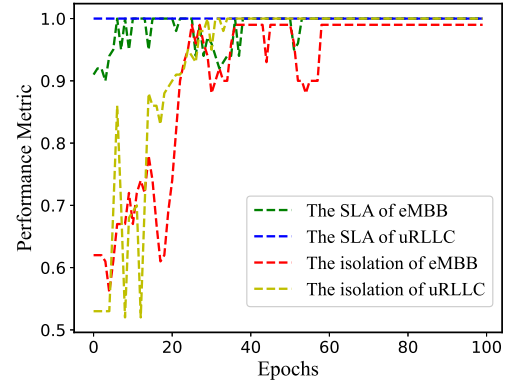}
\end{minipage}}

\subfigure[Hard-DQN]{
\label{iso2}
\begin{minipage}[t]{2.8in}
\centering
\includegraphics[width=2.8in]{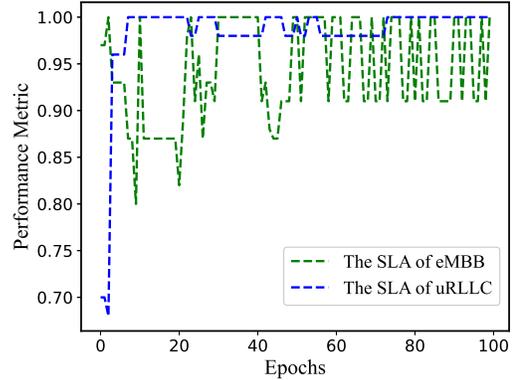}
\end{minipage}}
\caption{The SLA ration of convergence process for the proposed algorithm and Hard-DQN algorithm.}
\centering
\end{figure}
In a given area of $500 \times 500$m, one BS is located at the center with $43$dBm transmission power. $100$ RBs with each bandwidth $180$kHz are considered as total bandwidth resources. The pathloss model is consistent with \cite{myWCL2021}. Two slices corresponding to two types of services, i.e. eMBB and uRLLC services, are considered in the simulation. And the detailed slice parameters are summarized in Table \ref{tab:slice}. The values of $\gamma$, $\beta$ and $\rho$ are $0.9$, $5$ and $10$, respectively. The network architecture refers to DQN in \cite{DLMA}. The resource allocated to the common slice is 30 RBs in the initial phase, and the action set for one slice is $\{-5,-2,0,2,5\}$.
Three baseline algorithms are compared in our experiments:
\begin{itemize}
\item{\bf Optimal a Priori (OP):} Given a priori knowledge of traffic and SINR distributions of UEs, the optimal resource slicing is derived by exhaustive search. 
\item{\bf Hard-DQN}\cite{sun2019mix}: In this algorithm, a purely hard slicing framework using DQN is utilized.
\item{\bf NVS}\cite{kokku2011nvs}: NVS considers a static weight-based slicing with the assumption that the channel status of each user in the slice is known in priori.
\end{itemize}

\subsection{The Analysis of Convergence Process}
As Fig. \ref{fig:train} shows, the rewards of the proposed and the Hard-DQN algorithms are low initially and increase with training until they converge to the same level. It can be observed that the proposed algorithm converges slightly faster than the Hard-DQN algorithm. Since the setting of the common slice increases the SLA satisfaction on the exploration phase if compared with purely hard scheme.

Fig. \ref{iso1} and Fig. \ref{iso2} demonstrate the SLA satisfaction ratio of two algorithms in the first 100 epochs when training the agents. Observing Fig. \ref{iso1}, the SLA of uRLLC is always guaranteed. The reasons lie in two aspects. First, the packet size of the uRLLC slice is much smaller than the eMBB slice so that the required resource is lesser than the eMBB slice. Second, the shared resource of common slice prevents extreme scenarios, e.g. most RBs are allocated to eMBB slice. Similarly, the SLA of the eMBB slice is guaranteed after about 50 epochs. Compared with this, the uRLLC slice's SLA of the Hard-DQN algorithm can be guaranteed only after 70 epochs, and the eMBB SLA always fluctuates at the first 100 epochs. Naturally, the isolation degree of two slices of the proposed algorithm cannot approach the required thresholds at the initial phases and the isolation degree of Hard-DQN is always 1.
However, it is pointless to discuss isolation when the slices' SLA cannot be guaranteed. Furthermore, the isolation degree of the proposed algorithm can achieve the required thresholds after 60 epochs as shown in \ref{iso1}. 

\subsection{Performance Comparison}
\begin{figure}[!t]
    \centering
    \includegraphics[width=2.8in]{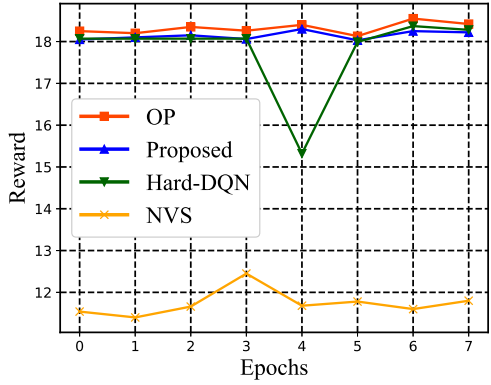}
    \caption{The rewards of four algorithms}
    \label{fig:reward_com}
\end{figure}

Fig. \ref{fig:reward_com} shows the achievable reward of four algorithms after two DQN-based algorithms converge. First, both the proposed algorithm and Hard-DQN can achieve approximately optimal performance. However, the performance of Hard-DQN fluctuates at $6$-th epoch due to a purely hard scheme. Second, the proposed algorithm far outperforms the NVS algorithm. Since NVS considers a static bandwidth provisioning slicing based on the aggregate throughput, it cannot satisfy the demand of mixed SLAs, e.g. latency and reliability metrics.
\section{Conclusion}
\label{sect:conc}
In this paper, we proposed a hard and soft hybrid slicing framework that introduces the common slice setting. A DRL-based solution is carefully designed. The comparison experiments indicate the proposed solution can guarantee slices' SLA all the time, even in the initial training phase. Moreover, it achieves near-optimal performance in terms of SLA satisfaction, spectrum efficiency and isolation.

\section*{Acknowledgement}
This work was supported in part by the National Natural Science Foundation of China (NSFC) under Grant 61871262, 62071284, and 61901251,  the National Key R\&D Program of China grants 2017YFE0121400 and 2019YFE0196600, the Innovation Program of Shanghai Municipal Science and Technology Commission grant 20JC1416400, Pudong New Area Science \& Technology Development Fund, Key-Area Research and Development Program of Guangdong Province grant 2020B0101130012, Foshan Science and Technology Innovation Team Project grant FS0AA-KJ919-4402-0060, and research funds from Shanghai Institute for Advanced Communication and Data Science (SICS).

\bibliographystyle{IEEEtran}
\bibliography{IEEEabrv,rf}

\end{document}